\documentclass[sigconf]{acmart}

\AtBeginDocument{%
  }

\setcopyright{none}
\copyrightyear{2025}
\acmYear{2025}
\acmDOI{XXXXXXX.XXXXXXX}

\usepackage[nolist]{acronym}
\begin{acronym}
    \acro{HCI}{Human-Computer Interaction}
    \acro{CSCW}{Computer-Supported Collaborative Work}
    \acro{CS}{Computer Science}
\end{acronym}
\usepackage{multirow}
\usepackage{subfigure} 
\usepackage{subcaption}
\usepackage{booktabs}
\usepackage{threeparttable}
\usepackage{siunitx}
\usepackage{adjustbox}
\usepackage{array}
\usepackage{threeparttable}
\usepackage{url}
\usepackage{hyperref}
\usepackage{_custom}
\usepackage{balance}

\begin{document}

\author{Yinuo Yang}
\authornote{Both authors contributed equally to this work.}
\affiliation{%
  \institution{University of Notre Dame}
  \city{Notre Dame}
  \state{IN}
  \country{USA}
}
\email{yinooyang@nd.edu}

\author{Zheng Zhang}
\authornotemark[1]
\affiliation{%
  \institution{University of Notre Dame}
  \city{Notre Dame}
  \state{IN}
  \country{USA}
}
\email{zzhang37@nd.edu}

\author{Ningzhi Tang}
\affiliation{%
  \institution{University of Notre Dame}
  \city{Notre Dame}
  \state{IN}
  \country{USA}
}
\email{ntang@nd.edu}

\author{Xu Wang}
\affiliation{%
  \institution{University of Michigan}
  \city{Ann Arbor}
  \state{MI}
  \country{USA}
}
\email{xwanghci@umich.edu}

\author{Alex Ambrose}
\affiliation{%
  \institution{University of Notre Dame}
  \city{Notre Dame}
  \state{IN}
  \country{USA}
}
\email{gambrose@nd.edu}

\author{Nathaniel Myers}
\affiliation{%
  \institution{University of Notre Dame}
  \city{Notre Dame}
  \state{IN}
  \country{USA}
}
\email{nmyers3@nd.edu}

\author{Patrick Clauss}
\affiliation{%
  \institution{University of Notre Dame}
  \city{Notre Dame}
  \state{IN}
  \country{USA}
}
\email{pclauss@nd.edu}

\author{Toby Jia-Jun Li}
\affiliation{%
  \institution{University of Notre Dame}
  \city{Notre Dame}
  \state{IN}
  \country{USA}
}
\email{toby.j.li@nd.edu}

\renewcommand{\shortauthors}{Yang and Zhang et al.}
\newcommand{\ino}[1]{\textcolor{purple}{\textbf{*Yinuo*}: #1}}

\begin{abstract}
AI-supported writing tools show strong potential for scaffolding students’ learning of argumentative writing. Prior work has demonstrated the benefits of AI-supported cognitive scaffolds, such as idea exploration and argument refinement, but how these features function in authentic classroom settings remains underexplored. In this paper, we investigate the classroom integration of an AI-supported writing tool, VISAR. We deployed VISAR in an undergraduate writing course across three sections for one week each over two semesters (49 students total). Using a mixed-methods approach that combines interaction logs, writing artifact analysis, surveys, and interviews, we examine how students used VISAR features in authentic writing tasks. Our findings confirm that students appropriated AI-supported cognitive scaffolds for writing learning and achieved measurable learning gains. While prior studies suggest that students may bypass important cognitive processes when using AI writing assistants, our classroom deployment shows that when systems provide structured supports for planning and targeted generation, students naturally choose to engage with these cognition-preserving scaffolds. These learning-oriented interaction patterns were positively associated with argumentative writing quality, improved conceptual understanding, and emerging critical AI literacy, highlighting the design value of cognition-preserving features in AI writing tools. Together, these findings provide empirical evidence of how AI-supported writing scaffolds operate in authentic classroom contexts and offer design insights for future learning-oriented AI writing tools.

\end{abstract}



\title{Lessons from Real-World Deployment of a Cognition-Preserving Writing Tool: Students Actively Engage with Critical Thinking and Planning Affordances}



\received{20 February 2007}
\received[revised]{12 March 2009}
\received[accepted]{5 June 2009}


\maketitle
\pagestyle{plain}

\section{Introduction}

Recent advances in generative AI have created new opportunities to support argumentative writing education~\cite{su2023collaborating, liu-august-2025-crafting}. A growing number of AI-based writing tools now provide assistance across multiple stages of the writing process. For example, by suggesting candidate claims or insightful questions during brainstorming, prompting reflection on logical weaknesses during revision, or offering alternative formulations of text~\cite{zhang2023visar, gero2022sparks, liu-august-2025-crafting}. These capabilities raise important questions for writing education, where the goal is not only to improve surface-level text quality, but to support students’ development of argumentative reasoning, rhetorical awareness, and metacognitive control over writing processes.

The question of whether and in what ways AI-supported writing tools enhance students' learning is still unanswered. Argumentative writing is a complex, process-oriented competence that requires sustained planning, monitoring, and revision of how claims, reasons, and evidence fit together~\cite{flowerHayes1981cognitive, zimmerman2002srl, winneHadwin1998studying}. Improvements in fluency or organization do not necessarily translate into deeper reasoning or transferable writing skills~\cite{wingate2012argument, andrews2010argumentation}. At the same time, generative AI introduces new risks for learning: students may over-rely on AI-generated content, outsource key reasoning decisions, or repeatedly regenerate drafts in ways that reduce engagement with argumentative construction~\cite{kosmyna2025brainchatgptaccumulationcognitive, kosmyna2025your, jin2025agencyGap, zare2025chatgptEssayRecall}. Moreover, unreliable outputs, such as hallucinated citations, require students to actively verify AI suggestions, making learning outcomes contingent not only on tool capabilities but on how students interact with them~\cite{kasneci2023chatgpt}. \looseness=-1

A growing body of research has examined AI-supported writing tools through system design, controlled experiments, and short-term evaluations~\cite{zhang2023visar, gero2022sparks, su2023collaborating, huang2023using, kasneci2023chatgpt, kosmyna2025your, jin2025agencyGap, zare2025chatgptEssayRecall}. Prior studies have shown that AI can support some stages of writing, that instructional framing influences how AI tools are used in educational settings~\cite{su2023collaborating, huang2023using}. Yet, much of this work treats ``AI support'' as a relatively unified category. Less is known about how different forms of AI interaction within the same tool function differently for learning, particularly under the constraints of real classroom settings where students work toward graded assignments, manage time pressure, and integrate AI into their own writing workflows.

In this paper, we argue that the key question is not whether AI-supported writing tools help students learn, but which forms of AI interaction support learning-oriented writing processes, and which may inadvertently undermine them. Addressing this question requires moving beyond outcome-only evaluations toward process-oriented analyses that examine how students actually engage with AI-supported features during authentic writing tasks. In-the-wild classroom deployments are therefore not merely confirmatory evaluations of tool effectiveness, but diagnostic lenses that can surface pedagogically consequential distinctions in how different AI-supported behaviors influence writing processes and learning-relevant practices.

To investigate these distinctions, we conducted an in-the-wild classroom deployment of VISAR~\cite{zhang2023visar}, a AI-supported argumentative writing tool that integrates multiple forms of AI assistance, such as idea elaboration, targeted argumentative scaffolds, visual planning, and draft regeneration \cite{blair2011groundwork, astute2023academic, ferretti2019argumentative, lovejoy2017great},within a single workflow. VISAR represents a category of tools designed specifically to promote learning goals in writing education, contrasting with approaches that integrate general-purpose AI tools (e.g., ChatGPT) into writing classrooms as task-agnostic support~\cite{su2023collaborating, huang2023using}. VISAR serves as a research probe allowing us to observe how students selectively appropriate different types of AI support when revising course assignments, rather than as a vehicle for introducing new system capabilities. We deployed VISAR in three sections over two semesters of an undergraduate writing course (N = 49) for one week per section and collected a rich dataset of interaction logs, writing artifacts, post-use surveys, and interviews with students and instructors.

Using a mixed-methods approach, we examine how students appropriated VISAR’s learning-oriented scaffolds in authentic writing workflows and how different patterns of interaction~\cite{kizilcec2013deconstructing} relate to argument quality, metacognitive regulation, and emerging forms of critical AI literacy.


Our findings confirm that students appropriated AI-supported cognitive scaffolds for writing learning and achieved measurable learning gains. While prior studies suggest that students may bypass important cognitive processes when using AI writing assistants~\cite{kosmyna2025brainchatgptaccumulationcognitive}, our classroom deployment shows that when systems provide structured supports for planning and targeted generation, students naturally choose to engage with these cognition-preserving scaffolds. This highlights the design value of embedding cognition-preserving features in AI writing tools to support learning-oriented appropriation.\looseness=-1

Consistent with these learning gains, our quantitative analyses reveal that students’ engagement in structured prompting and visual reasoning is positively associated with the quality of their argumentative writing. We also observe a significant increase in pre- to post-VISAR-use quiz scores, indicating improved conceptual understanding of core writing concepts. Furthermore, analysis of interaction logs reveals three distinct patterns of tool use, reflecting different appropriation strategies and workflows across students. These findings suggest that students do not engage with AI writing tools in a uniform way, but instead adopt diverse usage patterns aligned with their learning goals and writing practices.
Beyond writing quality, our qualitative analyses further suggest that students’ interactions with VISAR supported metacognitive regulation (e.g., planning, monitoring, and revising their writing strategies) and fostered emerging forms of critical AI literacy, such as evaluating the reliability of AI outputs and selectively integrating suggestions rather than accepting them uncritically.

This work makes three contributions.
\begin{itemize}
    \item First, we provide deployment-based evidence from authentic classroom use of a AI-supported writing tool, documenting how students appropriate multiple forms of AI assistance under real instructional constraints.
    \item Second, we offer a process-oriented account that links specific AI interaction behaviors, not AI use in general, to argument quality, metacognitive regulation, and critical AI literacy practices.
    \item Third, we derive design and pedagogical implications for integrating AI writing support.
\end{itemize}

\section{Related Work}

\subsection{Learning-Oriented AI Writing Support}
Generative AI systems such as ChatGPT are playing an increasingly important role in writing education~\cite{su2023collaborating}. Compared with earlier automated writing evaluation systems that primarily focused on surface-level features such as grammar and mechanics~\cite{attali2006automated, lewkow2016scalableAWE}, AI can offer richer forms of support, including proposing perspectives, questions, and candidate lines of argument, as well as enabling rapid iteration during writing~\cite{kasneci2023chatgpt, taecharungroj2023can}. Prior work has explored a range of AI-powered writing tools that support ideation and revision, including systems for rewriting or expansion~\cite{coenen2021wordcraft}, next-step suggestions~\cite{lee2022coauthor}, narrative ideation support~\cite{singh2023hide}, and structured prompting for writing tasks~\cite{gero2022sparks}. Empirical studies further show that writers often engage with AI suggestions in nuanced ways, selectively adapting generated content rather than directly copying it~\cite{bhat2023interacting}.\looseness=-1

At the same time, whether AI-supported writing tools contribute to students’ \emph{learning} remains an open question. Prior work has highlighted risks such as over-reliance, fabricated references, and cognitive offloading~\cite{kosmyna2025brainchatgptaccumulationcognitive, ariyaratne2023comparison, altmae2023artificial, kosmyna2025your}. Education-focused studies suggest that heavy reliance on ChatGPT may undermine learners’ independent writing capabilities over time~\cite{zare2025chatgptEssayRecall}, and that learners’ AI literacy can influence writing performance once AI support is removed~\cite{jin2025agencyGap}. Complementing these accounts, factors such as trust and satisfaction have been shown to influence over- versus appropriate reliance on AI assistance~\cite{pitts2025students}. Together, these findings reinforce long-standing arguments from writing pedagogy that effective writing support should scaffold thinking and reasoning processes, rather than merely improve final text quality~\cite{liu-august-2025-crafting}.

\subsection{Argumentation Learning and Writing Processes}
Argumentative writing requires coordinating claims, reasons, evidence, and counterarguments, as well as continuously monitoring and revising the evolving argument as a whole~\cite{toulmin2003uses, flowerHayes1981cognitive}. Educational research has long shown that students often struggle with these demands, including generating viable lines of argument, selecting and integrating evidence, and maintaining coherence across sections~\cite{wingate2012argument, andrews2010argumentation}. From a learning perspective, these challenges are closely tied to metacognitive regulation in writing, commonly characterized in terms of planning, monitoring, and revision~\cite{zimmerman2002srl, winneHadwin1998studying, schrawDennison1994MAI}.

Accordingly, learning-oriented writing supports have explored both representational and prompt-based scaffolds to externalize and guide argumentative reasoning. For example, argument mapping and other structural representations can make relationships among argumentative components more inspectable and support reflection on overall coherence~\cite{rapanta2016use, cullen2018improving, van2015using}. More recently, LLM-based approaches have been used to provide targeted critique, alternative perspectives, or guided prompts during writing~\cite{gero2022sparks, zhang2023visar}. While these approaches highlight the potential of scaffolding argumentative reasoning, prior work has primarily examined the design of such supports or their effects in controlled or short-term settings, leaving open questions about how students integrate multiple forms of support over time in authentic classroom workflows.

\subsection{Classroom Deployment and Process-Oriented Evaluation}
Across prior work on AI writing support and argumentation learning, a key gap is the lack of deployment-based evidence that connects \emph{writing processes} to \emph{learning-relevant outcomes} in real classroom contexts. Conceptual and pedagogical discussions emphasize the importance of preserving student agency and fostering critical engagement when integrating AI into writing instruction~\cite{su2023collaborating, huang2023using, leander2020critical}. However, classroom settings introduce sociotechnical constraints, such as assignment goals, time pressure, and instructional framing, that influence how tools are actually used. As a result, learning-relevant effects may be reflected not only in final writing quality, but also in process indicators such as cycles of planning and revision, selective uptake of scaffolds, and verification of AI suggestions~\cite{flowerHayes1981cognitive,zimmerman2002srl,schrawDennison1994MAI,lu2024generative}.

Process-oriented evaluation approaches address this gap by examining learners’ interactions and writing artifacts alongside outcome measures and self-reports. Such approaches make it possible to characterize heterogeneous engagement patterns, relate concrete writing behaviors to dimensions of argument quality, and interpret these patterns in light of students’ goals, perceptions, and instructional context~\cite{lewkow2016scalableAWE, winneHadwin1998studying,lu2023readingquizmaker}. 

Recent work have increasingly adopted process-oriented evaluation frameworks that combine interaction traces, learning artifacts, and qualitative accounts to examine how users appropriate AI- or visualization-mediated supports in practice~\cite{yang2025spark,10.1145/3746059.3747742,zhang2026editrail,ai2025nli4volvis,ai2025evaluation,yang2024vizcode,ai2025keo}. Building on this perspective, our work reports an in-the-wild classroom deployment study that uses process-oriented evidence to examine AI-supported argumentative writing as a learning and metacognitive process. Rather than introducing new system capabilities, we focus on how students appropriate multiple forms of AI support under authentic classroom constraints, and how distinct process-level behaviors relate to argument quality and learning-relevant practices.

\section{Method}

\subsection{Instructional Context and Participants}
We conducted a deployment study of an AI argumentative writing tool in a core first-year \textit{Writing and Rhetoric} course, designed to develop foundational skills in argumentation, rhetorical reasoning, and effective communication, at a large, private, R1-classified university in the Midwestern United States\footnote{The study protocol was reviewed and approved by the IRB at our institution.}.

The curriculum emphasizes thesis formation, claim structuring, and evidence usage. The study spanned three sections of the same course over two semesters, taught by two instructors with  expertise in writing pedagogy, digital rhetorics and generative AI in writing education. Each section completed a one-week deployment (roughly two-thirds into the semester) of an AI-supported writing tool featuring idea exploration and visual planning. 

In total, 49 undergraduate students participated across the three sections. Background surveys completed by 32 students indicated that the majority (19) self-identified as intermediate writers, with an additional seven reporting advanced experience. This distribution suggests a functional baseline in academic writing. Students reported common planning strategies such as outlining and free-writing, while citing specific challenges in organizing ideas and connecting evidence—cognitive. Together, the instructional context and participants provide an ecologically valid setting for examining how students appropriate AI-supported writing scaffolds under authentic classroom constraints.



\vspace{-0.5em}
\subsection{Study Procedure}
The study followed a consistent deployment protocol across three sections, integrating the AI-supported writing tool VISAR~\cite{zhang2023visar} into the standard course workflow for one week per section. While the core deployment procedure was consistent across sections, one section (Sec. 2, $N=14$) additionally incorporated a brief pre- and post-session quiz to assess potential short-term gains in argumentative structuring and rhetorical reasoning.

The deployment took place after students had completed an initial draft of an open-topic argumentative essay. Students were required to select a personally meaningful topic ranging from campus policies to broader social or cultural issues, and develop an argument around it\footnote{Essay instruction is provided in the supplementary material: \href{https://osf.io/wsdnb/overview?view_only=476a2071749f47f9bb9d9c98342b81ae}{OSF link}.}. This open-ended assignment encouraged an authentic authorial voice in the writing process.
Each deployment began with a 75-minute in-class session in which the instructor introduced VISAR and demonstrated its core functionalities. Students then used the tool autonomously in class to reflect on and revise their essays, focusing on idea exploration, argument refinement supported by LLM-based feedback, and visual planning of argument structure. They were encouraged to integrate VISAR into their workflows in ways that aligned with their individual writing practices, using the system as a flexible writing support tool. After the deployment, students were invited to complete a post-use questionnaire. We also conducted semi-structured interviews with four students and two instructors to gain deeper insights into their experiences with VISAR.

\vspace{-0.5em}
\subsection{The VISAR System}
VISAR~\cite{zhang2023visar} is a AI-supported argumentative writing tool designed specifically for writing education. It integrates a visual argument-planning workspace with targeted AI scaffolds that support core practices of argumentation (e.g., articulating claims) and the metacognitive regulation of writing (planning, monitoring, and revision).

\textbf{Visual planning workspace.} VISAR provides a node-and-edge canvas where students externalize the structure of their essays. Nodes represent argumentative units (e.g., claim, evidence), and links represent relationships among them. This workspace is intended to make otherwise implicit reasoning structures visible and revisable during drafting and revision.

The visual workspace is bi-directionally synchronized with the text editor. Edits made in the text editor are reflected in the visual representation, and students can also manipulate the visual structure (e.g., add notes, rearrange nodes, or change connections) with corresponding changes propagated back to the text editor.

\textbf{AI-supported scaffolds and generation.} VISAR offers multiple forms of AI support: \textbf{(1) Elaboration:} hierarchical prompts that help students expand and refine lines of argument.
\textbf{(2) Argumentative Sparks:} targeted scaffolds that prompt evidence suggestions, counterarguments, and detection of potential logical weaknesses.
\textbf{(3) LLM Draft Regeneration:} generation of \emph{intentionally low-fidelity} draft text that students can revise. These drafts are designed to function as concrete representations that help students rapidly prototype alternative argumentative strategies and structures and observe how they can be operationalized in text. 
\textbf{(4) Synchronized Visual--Text Editing:} edits in the visual plan can be used to inform drafting, and drafting can surface structures for further visual inspection.

In this study, we use VISAR as a research probe to examine how students selectively appropriate these different forms of support under authentic classroom constraints.

\vspace{-0.5em}
\subsection{Data Collection}

\textit{System Interaction Logging.}
VISAR recorded detailed user interaction data throughout the deployment by using an artifact history logging feature that captured writing states at finer granularity. Each AI-assisted generation or revision triggered a snapshot of the updated text and visual outline. Additionally, for manual writing processes, the system automatically saved a snapshot every 60 seconds. This enabled temporal analysis of the writing process, including pacing, sequencing, and iterative refinement strategies.

\textit{Writing Artifacts.}
We collected all intermediate and final versions of essays produced in VISAR, including visual outlines, AI-generated paragraphs, and student-authored revisions. These artifacts served as essential data sources for analyzing how students translated planning artifacts into structured argumentative writing.

\textit{After-Use Questionnaire and Interviews.}
After the deployment, 34 students completed the after-use questionnaire reflecting on their experiences with VISAR.
Additionally, we conducted semi-structured interviews with four student participants and with two course instructors. The interviews probed their experiences using VISAR and attitudes toward AI-assisted writing support. 

\vspace{-0.5em}
\subsection{Data Analysis}

\subsubsection{Student Engagement and Use Patterns}
\label{sec:cluster_analysis}

Identifying engagement archetypes through behavioral clustering has been widely used in learning analytics to characterize heterogeneous learner trajectories~\cite{kizilcec2013deconstructing}.
To characterize how students engaged with VISAR during authentic writing tasks, we conducted a fine-grained analysis of system interaction logs. This analysis focused on quantifying students’ use of different system components, with the goal of understanding how learners navigated VISAR’s features and which forms of support were most frequently appropriated during writing.

We operationalized students’ interaction behaviors by categorizing logged events into four functional metrics, each corresponding to a distinct form of system-supported activity. These metrics capture both AI-mediated and user-driven interactions and allow us to examine patterns of engagement, relative reliance on different scaffolds, and changes in interaction over time.

The behaviors were categorized into four metrics:
\textbf{(1) Elaboration:} Usage of AI hierarchical writing ideation suggestions to brainstorm argumentative structure. 
\textbf{(2) Spark Usage:} Activation of AI argumentative scaffolds, including evidence suggestions, counterarguments, and logical weakness detection. 
\textbf{(3) LLM Draft Regeneration:} User-initiated requests for AI-supported regeneration of draft content. 
\textbf{(4) Visual Interaction:} Manipulations of the argument map, including node/edge creation, deletion, edition and hierarchy adjustments.

Based on the interaction behavior data, we performed a $k$-means clustering analysis~\cite{jain2010data} to uncover distinct patterns of tool usage among students. The clustering model utilized four operational metrics explained above. Feature vectors were standardized using $z$-scores to account for scale disparities~\cite{james2013introduction}. The optimal number of clusters was determined via the elbow method to balance model fit with interpretability~\cite{thorndike1953belongs}.


\subsubsection{Learning Outcomes and Preceived Utility}
\label{sec:outcome_analysis}

\paragraph{Assessment of Argument Quality.}
Using student essays collected with VISAR during classroom writing sessions, two researchers independently screened the essays to determine whether each constituted a substantive argumentative essay suitable for analysis. Essays were labeled as either \textit{retained} or \textit{excluded} based on whether they demonstrated meaningful argumentative content, as opposed to being incomplete, off-topic, or otherwise non-substantive. Inter-rater agreement was assessed to ensure consistency in screening decisions. We computed Cohen’s $\kappa$ to quantify inter-rater reliability and observed substantial agreement ($\kappa = 0.84$)~\cite{mchugh2012interrater}. Disagreements were resolved through discussion to reach consensus. Only essays meeting the inclusion criteria were retained for subsequent analyses.\looseness=-1

To assess the quality of arguments, an external expert in English writing education was hired to evaluate all the retained essays. The expert has 10+ years of experience teaching academic and exam-oriented English writing and has taught hundreds of students for standardized test in English writing e.g., the IELTS exam. Each essay was scored along five established dimensions of argumentative quality based on Toulmin model~\cite{toulmin2003uses}: \textit{(1) Claim Articulation}, \textit{(2) Grounds (Supporting Evidence)}, \textit{(3) Warrant and Causal Logic}, \textit{(4) Counter-Consideration and Qualification}, and \textit{(5) Argument Progression Across Sections}\footnote{Grading rubric and sample are provided in the supplementary material: \href{https://osf.io/wsdnb/overview?view_only=476a2071749f47f9bb9d9c98342b81ae}{OSF link}.}.


Based on these scores, we conducted multiple linear regression analyses to examine associations between interaction behaviors and argument quality outcomes~\cite{montgomery2012applied}. Each argument quality dimension was modeled as a separate outcome variable. For each model, four interaction behavior metrics—\textit{Elaboration}, \textit{Spark Usage}, \textit{LLM Draft Regeneration}, and \textit{Visual Interactions}—were entered simultaneously as predictors to account for their relative contributions. All predictor variables were standardized prior to analysis to facilitate comparison of effect sizes across interaction types~\cite{cohen1988statistical}. Model fit was assessed using standard goodness-of-fit measures, and all models were estimated on the same set of retained essays.


\textit{Quiz Performance and Conceptual Learning.}
To assess students’ understanding of argumentative structures and rhetorical strategies, including claim formation, evidence integration, and counter-argumentation, we analyzed students’ performance on a short quiz administered before and after the classroom deployment of VISAR. This analysis focused on students in Sec.~2, for whom both pre- and post-deployment assessments were available.

The quiz\footnote{The full quiz is provided in the supplementary material: \href{https://osf.io/wsdnb/overview?view_only=476a2071749f47f9bb9d9c98342b81ae}{OSF link}.} consisted of eight items designed by the course instructors to probe key concepts in argumentative writing. These items covered core concepts in argumentative writing, including the relationship between topic sentences and thesis statements, the role of evidence in supporting claims, paragraph-level organization, and the structural requirements of counterclaims and rebuttals.

We employed a within-subject design in which students completed the same quiz before and after the classroom deployment of VISAR, enabling aggregate- and item-level comparisons of performance across time. Quiz scores were treated as paired observations, and pre–post differences were analyzed using paired $t$-tests~\cite{student1908probable}. We additionally conducted item-level analyses to examine changes in responses to individual questions, allowing us to assess shifts in students’ demonstrated understanding of specific argumentative concepts, including claim–thesis alignment, evidence use, and the structural role of counterclaims and rebuttals.



\textit{Perceived Utility and Workflow Integration. }
We analyzed post-use questionnaire data to characterize students’ perceptions of VISAR following the classroom deployment. Participation in the questionnaire was voluntary. The survey captured students’ self-reported experiences with the system, including perceived helpfulness of individual features, perceived impacts on writing quality, and how VISAR was incorporated into their writing workflows. Responses were summarized to identify overall patterns in students’ attitudes toward different aspects of the system and to contextualize findings from interaction log and interview analyses.

\subsubsection{Qualitative Data Analysis}
\label{sec:interview_analysis}
Following established open-coding protocols~\cite{braun2006using, lazar2017research}, two researchers jointly reviewed all interview transcripts and conducted open coding to identify recurring patterns related to system use, perceived value, and classroom integration. Through constant comparison and discussion, codes were refined and consolidated into a collaboratively developed codebook. The researchers then independently applied the finalized codebook to the full dataset, with discrepancies discussed and resolved through consensus to strengthen interpretive consistency.
This analysis examined how VISAR was appropriated in practice and how its affordances were interpreted.

\section{Quantitative Results}

\subsection{Student Engagement and Use Patterns}

\subsubsection{Overview of Patterns} \label{interaction_patterns}

We received 1,543 discrete interaction log events from 66 writing sessions ($N=49$ students). On average, students generated 23.4 interactions per session ($SD = 36.8$). 

\begin{table}[t]
\centering
\small
\setlength{\tabcolsep}{8pt}
\begin{tabular}{p{2.2cm} p{2.8cm} r r}
\hline
Category & Subcategory & Count & \% \\
\hline
Elaborate & -- & 163 & 10.34 \\
\hline
\multirow{3}{*}{Spark Usage} 
& Supporting Evidence & 106 & 6.73 \\
& Logical Weakness & 45 & 2.86 \\
& Counterargument & 51 & 3.24 \\
& \textit{Total} & \textit{202} & \textit{12.82} \\
\hline
LLM Draft Regenerate & -- & 95 & 6.03 \\
\hline
\multirow{3}{*}{Visual Interaction} 
& Node Create/Delete & 57 & 3.62 \\
& Node Graph Update & 61 & 3.87 \\
& Node Text Update & 998 & 63.32 \\
& \textit{Total} & \textit{1116} & \textit{70.81} \\
\hline
\textit{Overall Total} &  & \textit{1576} & \textit{100} \\
\hline
\end{tabular}
\caption{Distribution of interaction events across VISAR features (N = 1,543). Percentages are calculated over the full dataset and rounded to two decimals.}
    \vspace{-1cm}

\label{tab:interaction-distribution}
\end{table}

As detailed in Tab.~\ref{tab:interaction-distribution}, interaction activity was heavily concentrated within the visual workspace. Visual Interactions constituted the predominant mode of engagement (70.81\%, $n=1,116$), with the vast majority consisting of text updates within nodes (63.30\%). This indicates that students frequently composed and refined content directly within the visual structure.

Scaffolding features also saw significant usage. Spark Usage accounted for 12.82\% of total events ($n=202$), driven primarily by Evidence suggestions (6.73\%), followed by Counterarguments (3.24\%) and Logical Weakness detection (2.86\%). Additionally, Elaboration represented 10.34\% of interactions, while LLM Draft Regeneration accounted for 6.03\%.

The high frequency of visual planning and brainstorm-oriented interactions suggests that students actively leveraged VISAR’s scaffolding mechanisms to structure and refine their reasoning. The distribution of events across visual and textual components further indicates that students engaged in a recursive, multi-level authoring process that interleaved planning, scaffolding, and drafting.


\subsubsection{Cluster-Based Analysis of Engagement Archetypes}
\begin{figure}[t]
    \centering
    \includegraphics[width=0.8\linewidth]{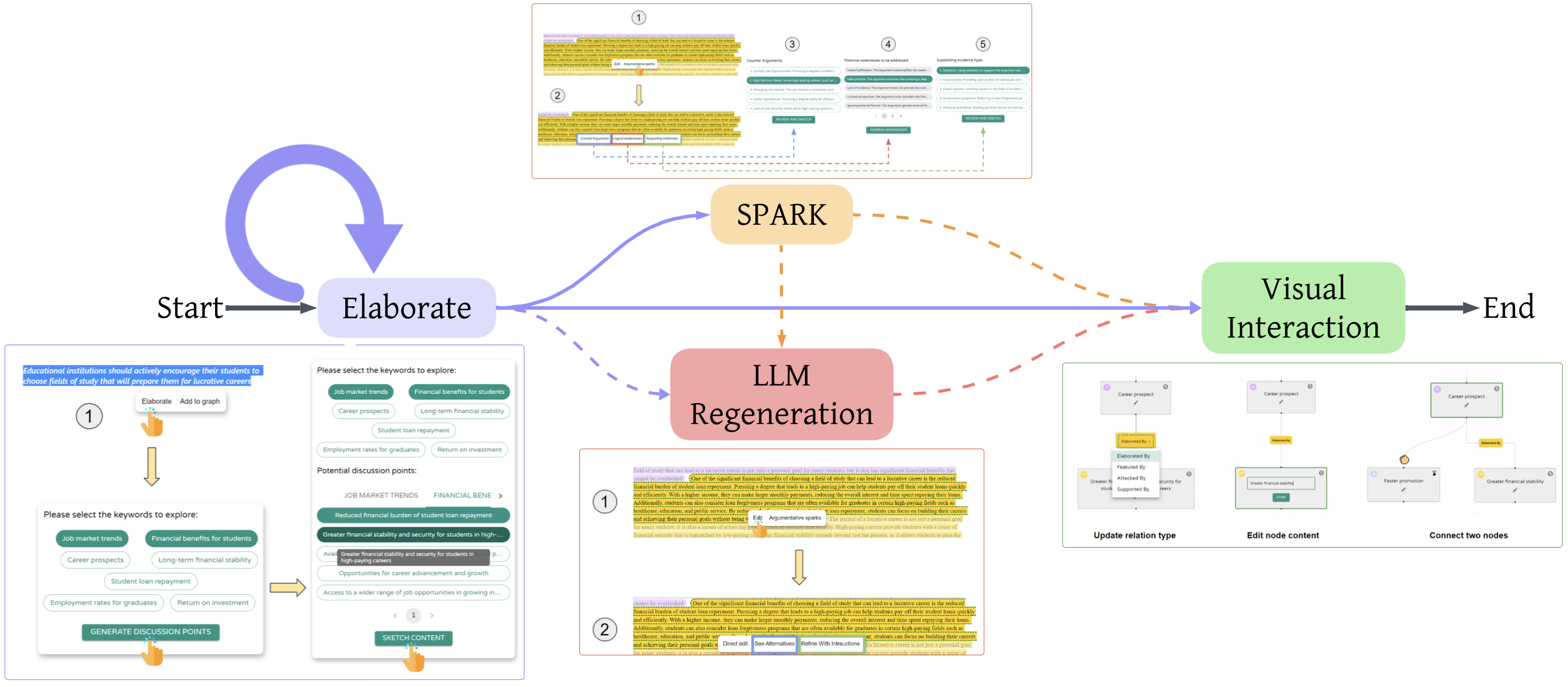}
    \par\smallskip
    {\small (a) Brainstorm-Oriented Explorers (Cluster 1)}
    \vspace{0em}
\includegraphics[width=0.8\linewidth]{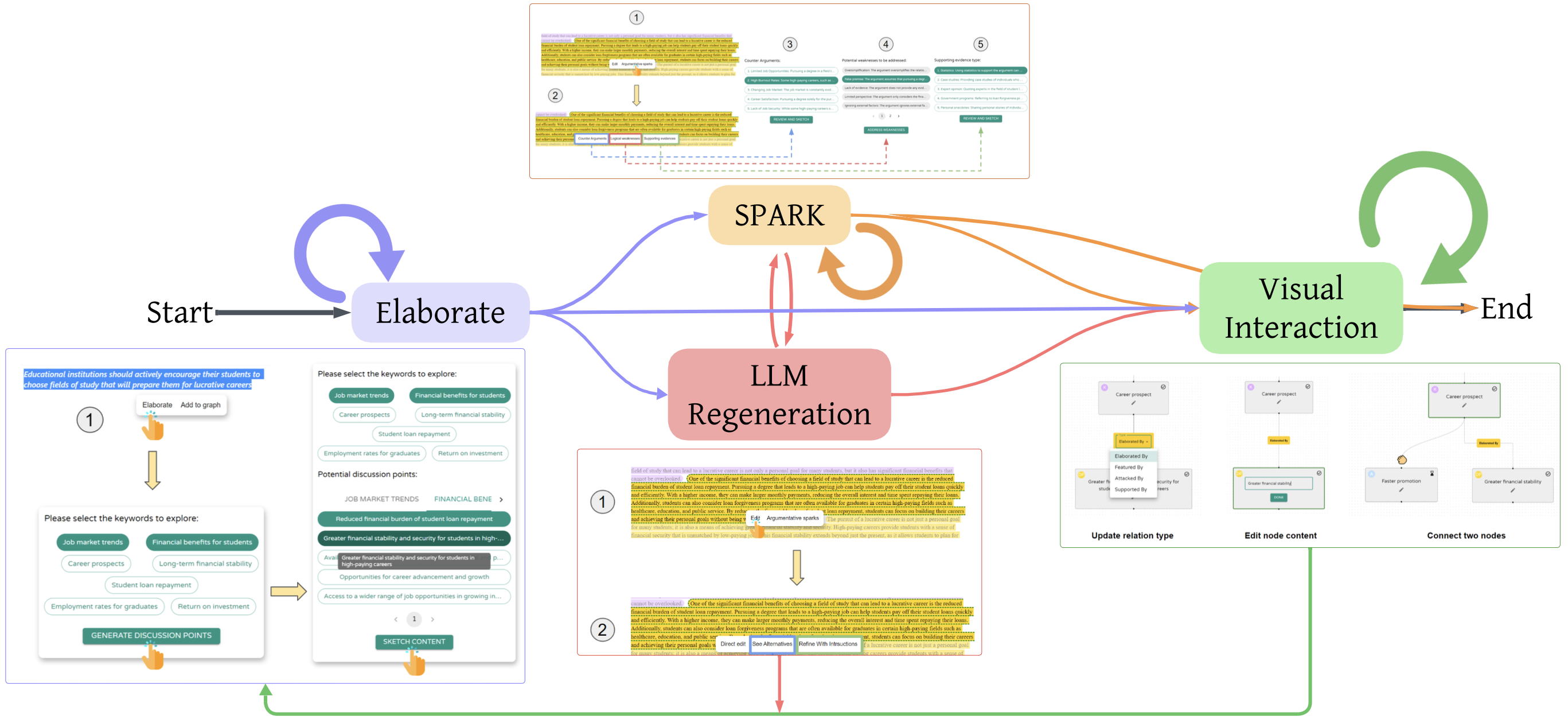}
    \par\smallskip
    {\small (b) Strategic Scaffolders (Cluster 2)}
    \vspace{0em}
\includegraphics[width=0.8\linewidth]{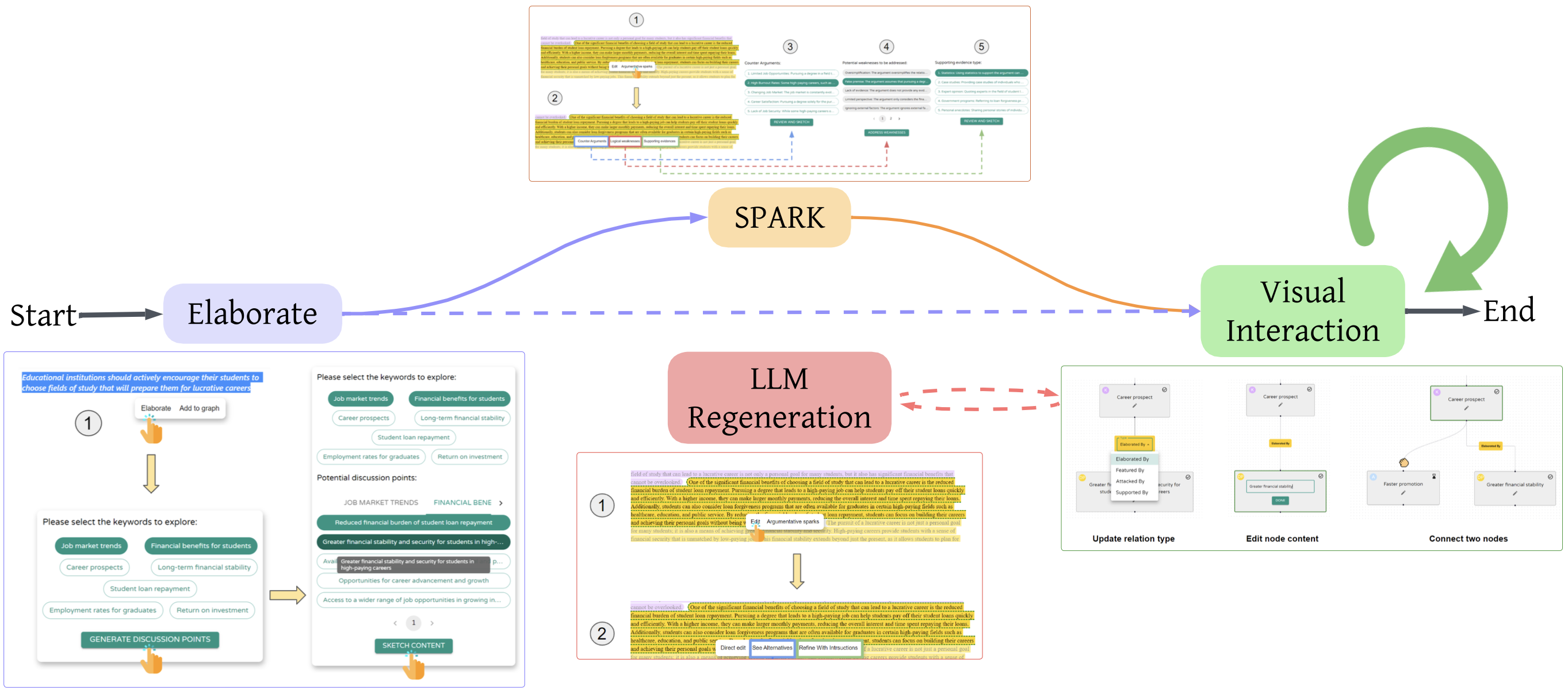}
    \par\smallskip
    {\small (c) Visual Planners (Cluster 3)}
    \caption{Feature usage profiles for the three engagement clusters identified through
    $k$-means clustering.}
    \vspace{-0.8cm}
    \label{fig:clusters}
\end{figure}

Following the $k$-means clustering method described in Sec.~\ref{sec:cluster_analysis}, we revealed three distinct engagement archetypes across 66 writing sessions, as visualized in Fig.~\ref{fig:clusters}:

\textbf{Cluster 1: Brainstorm-Oriented Explorers (N = 40).}
This largest group engaged with the system in a lightweight, brainstorm-focused manner. Students in this cluster made limited use of visual planning and drafting features but frequently interacted with elaboration prompts to clarify directions and explore argumentative possibilities. Their interaction patterns suggest that VISAR was primarily used as a conceptual support tool for brainstorming and planning, rather than as a primary drafting workspace.

\textbf{Cluster 2: Strategic Scaffolders (N = 22).}
Students in this cluster made frequent, sustained use of multiple system features, including the visual planning board, spark-based argumentative scaffolds, and LLM-assisted draft generation. Their sessions reflected an integrated workflow in which students organized ideas, solicited targeted feedback on evidence and logic, and iteratively refined their drafts. This pattern suggests strategic use of VISAR as a comprehensive scaffolding environment rather than a single-purpose writing aid.

\textbf{Cluster 3: Visual Planners (N = 4).}
A small subset of sessions exhibited exceptionally high levels of visual planning activity, far exceeding those in other clusters. These students repeatedly manipulated and revised visual argument structures while making comparatively little use of other features. This pattern indicates a strong preference for externalizing and refining argumentative structure through visual representations.

\vspace{-0.5cm}
\subsection{Learning Outcomes and Perceived Utility}
\subsubsection{Assessment of Argument Quality}

Based on the method described in Sec.~\ref{sec:outcome_analysis}, of the 66 essays collected, 37 were retained for analysis as substantive argumentative essays. The retained essays exhibited variation in length and level of development, with an average length of 709.65 words (SD = 513.26).

Across the five score dimensions, interaction behaviors with VISAR showed differentiated associations with argument quality. Regression analyses relating students’ interaction patterns to argument quality dimensions revealed distinct relationships between specific forms of system use (the detailed results are reported in Tab.~\ref{tab:reg_quality}).
Across models, we observe a clear and consistent pattern differentiating generative AI-driven behaviors from interactive, structure-oriented actions. \textit{Spark Usage} shows a strong positive association with several core dimensions of argument quality. In particular, higher \textit{Spark Usage} significantly predicts stronger \textit{Claim Articulation} ($\beta = 0.880$, $p < .001$) and \textit{Grounds} ($\beta = 0.435$, $p < .01$), as well as improved \textit{Argument Progression Across Sections} ($\beta = 0.640$, $p < .001$; Tab.~\ref{tab:reg_quality}, Columns~1,~2, and~5). These results suggest that engaging with Sparks, which was designed to prompt targeted reasoning and idea development, supports students in formulating clearer claims, grounding them with evidence, and maintaining coherence across sections of an essay.

In contrast, \textit{LLM Draft Regeneration} is consistently associated with lower argument quality across multiple dimensions when controlling for other behaviors. Greater reliance on \textit{LLM Draft Regeneration} predicts significantly lower scores in \textit{Claim Articulation} ($\beta = -0.896$, $p < .01$), \textit{Grounds} ($\beta = -0.321$, $p < .05$), \textit{Warrant and Causal Logic} ($\beta = -0.471$, $p < .05$), and \textit{Argument Progression Across Sections} ($\beta = -0.681$, $p < .01$; Tab.~\ref{tab:reg_quality}). This pattern indicates that repeatedly regenerating drafts may substitute for students’ own reasoning work, weakening the explicit articulation of claims, causal links, and overall argumentative coherence.

\textit{Visual Interactions} exhibit a complementary and more localized effect. While not predictive of claim clarity or overall progression, visual planning significantly predicts higher-quality \textit{Warrant and Causal Logic} ($\beta = 0.574$, $p < .01$) and \textit{Counter-Consideration and Qualification} ($\beta = 0.792$, $p < .01$; Tab.~\ref{tab:reg_quality}). These results suggest that visually organizing arguments, such as mapping relationships between claims, evidence, and counterarguments, particularly supports deeper reasoning processes that require making implicit assumptions explicit and engaging with alternative perspectives.\looseness=-1

Finally, \textit{Elaboration} does not emerge as a significant predictor in any model after accounting for other interaction behaviors (all $p > .1$), suggesting that surface-level expansion of text alone is insufficient to explain variation in argument quality when more targeted, reasoning-oriented interactions are considered.

Overall, the models explain a substantial proportion of variance across argument quality dimensions ($R^2 = 0.296$--$0.463$; Tab.~\ref{tab:reg_quality}), highlighting distinct and theoretically meaningful roles of different interaction behaviors. Taken together, these findings indicate that how students interact with AI-supported writing tools has important implications for the quality of their argumentative writing.

\begin{table*}[t]
\small
\centering

\begin{adjustbox}{max width=\textwidth}
\renewcommand{\arraystretch}{1.15}
\begin{tabular}{lccccc}
\toprule
& \textbf{(1) Claim} & \textbf{(2) Grounds} & \textbf{(3) Warrant} & \textbf{(4) Counter} & \textbf{(5) Progression} \\
\midrule
\textbf{Intercept} 
& 3.946 (0.177)*** 
& 4.568 (0.106)*** 
& 4.189 (0.146)*** 
& 2.622 (0.198)*** 
& 3.514 (0.155)*** \\

Elaboration 
& $-0.020$ (0.200) 
& $-0.007$ (0.119) 
& $-0.178$ (0.164) 
& $-0.289$ (0.223) 
& $-0.185$ (0.175) \\

Spark Usage 
& 0.880 (0.200)*** 
& 0.435 (0.119)** 
& 0.236 (0.165) 
& 0.093 (0.224) 
& 0.640 (0.176)*** \\

LLM Draft Regeneration 
& $-0.896$ (0.253)** 
& $-0.321$ (0.151)* 
& $-0.471$ (0.208)* 
& $-0.048$ (0.283) 
& $-0.681$ (0.222)** \\

Visual Interactions 
& 0.149 (0.255) 
& 0.091 (0.152) 
& 0.574 (0.210)** 
& 0.792 (0.285)** 
& $-0.187$ (0.224) \\

\midrule
$R^2$ & 0.453 & 0.308 & 0.296 & 0.412 & 0.463 \\
\bottomrule
\end{tabular}
\end{adjustbox}

\caption{Multiple linear regressions predicting argument quality dimensions from interaction behaviors (standardized predictors). Entries are standardized coefficients $\beta$ with standard errors in parentheses. Predictors were z-scored prior to analysis; intercepts are in outcome units.
* $p<.05$, ** $p<.01$, *** $p<.001$.
}
\vspace{-0.8cm}
\label{tab:reg_quality}
\end{table*}

\vspace{-0.5em}
\subsubsection{Quiz Performance and Conceptual Learning}
Based on the methods described in Sec.~\ref{sec:outcome_analysis}, among the 14 students in Sec.~2 who completed both the pre- and post-quizzes, we observed a substantial improvement in conceptual understanding, as reflected by a significant increase in scores ($p < .001$). The mean score increased from $5.79$ ($SD = 0.97$) on the pre-test to $7.50$ ($SD = 0.85$) on the post-test.
An item-level analysis identified marked improvements across several specific dimensions:
\begin{itemize}

\item\textbf{Evidence-Based Reasoning ($p < .05$):} There was an increase in accuracy regarding the requirement that every main claim must be supported by evidence. This suggests that VISAR's visual mapping and prompt-based scaffolds effectively reinforced the necessity of grounding arguments in empirical or logical support, moving students toward more rigorous evidence-based writing.
\item \textbf{Counter-argumentation Strategies ($p < .01$):} Students performed significantly better on advanced rhetorical items, such as the necessity of dedicated paragraphs for rebuttals and the strategic timing of counterclaims. This indicates that the ``Argumentative Sparks'' provided by the system helped students internalize complex strategies for addressing opposing viewpoints.
\item \textbf{Evidence Closure ($p < .0001$):} Students demonstrated a firmer grasp of fundamental writing rules, specifically the requirement that every main claim must be supported by evidence and the importance of structural ``links'' or ``lead-outs'' at the end of body paragraphs.
\end{itemize}

In summary, the elevation in quiz scores suggests that the deployment of VISAR helped students improve their metacognitive understanding of argumentative forms and structural logic.

\subsubsection{Perceived Utility and Workflow Integration}


\begin{figure*}[t]
    \centering
    \includegraphics[width=0.8\textwidth]{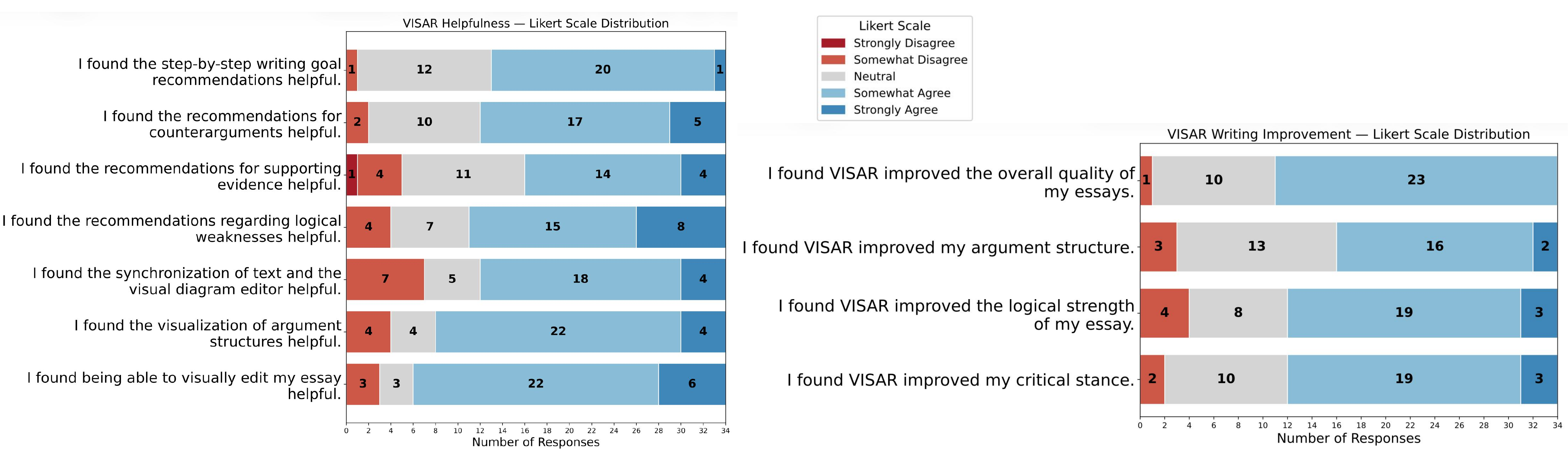}
    \caption{Students’ perceived helpfulness of different VISAR features and impact of VISAR on different aspects of writing quality on a five-point Likert scale.}
            \vspace{-0.2cm}

    \label{fig:helpfulness}
    \label{fig:improvement}
\end{figure*}

\begin{figure*}[t]
    \centering
    \includegraphics[width=0.8\textwidth]{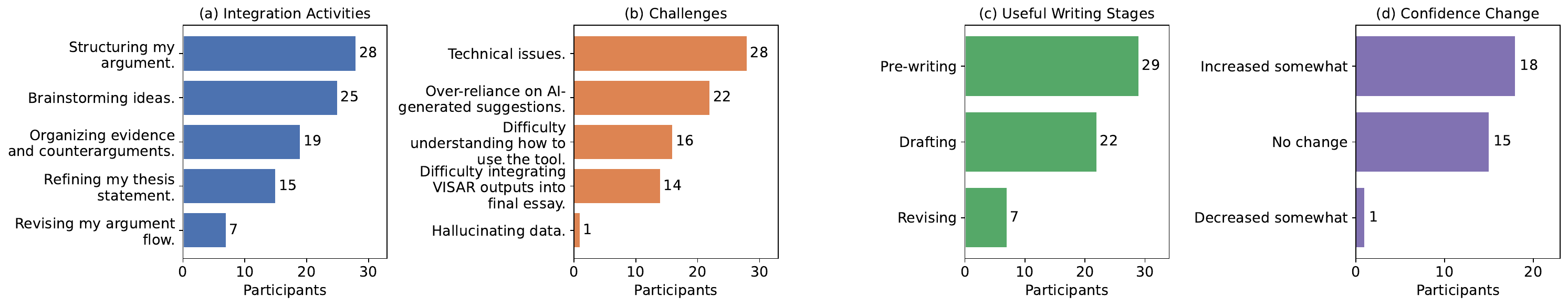}
    \caption{Survey results on how students integrated VISAR into their writing process: 
    (a) writing activities supported by VISAR, 
    (b) challenges encountered, 
    (c) writing stages where VISAR was most useful, and 
    (d) changes in writing confidence.}
        \vspace{-0.2cm}

    \label{fig:survey}
\end{figure*}

We received 34 responses from 49 students. Fig.~\ref{fig:helpfulness} summarizes students’ perceived helpfulness of VISAR features
on a five-point Likert scale. Overall, students reported positive perceptions across most
components, particularly for features supporting visual reasoning and argument structure.
A majority of participants agreed or strongly agreed that visually editing their essays and
viewing argument structures were helpful, as were the synchronized visual–text editing
capabilities. Scaffolded recommendations, including evidence suggestions, counterarguments,
and logical weakness detection, were also generally perceived as helpful. Step-by-step writing elaboration
recommendations received relatively fewer strong endorsements, suggesting that students
may have varied in how much they relied on higher-level procedural guidance. 

Students also reported perceived improvements in multiple dimensions of writing quality
(Fig.~\ref{fig:improvement}). Most participants indicated that VISAR helped improve the
logical strength and structural organization of their essays, as well as their overall essay
quality. Improvements in maintaining a critical stance were also reported. 

Beyond perceived outcomes, we further examined how students integrated VISAR into their
writing workflow and the challenges they encountered (Fig.~\ref{fig:survey}).
Most students reported using VISAR during pre-writing and drafting stages, with fewer
students relying on it primarily during later-stage revision. Integration activities most
commonly included structuring arguments and brainstorming ideas, followed by organizing
evidence and counterarguments. At the same time, students reported several challenges in integrating VISAR outputs into their final essays.
Concerns about over-reliance on AI-generated suggestions were also noted by a subset of
participants, while hallucinated content was rarely reported. Finally, most students
reported either stable or slightly increased confidence in their writing, suggesting that
VISAR did not undermine students’ sense of authorship or self-efficacy.\looseness=1

\vspace{-1em}
\section{Qualitative Findings}

\subsection{Supporting Idea Exploration During Pre-Drafting}

Across the classroom deployment, students frequently engaged with AI-supported idea elaboration during pre-drafting, when they were transitioning from open-ended brainstorming to more structured exploration of argumentative possibilities. Many participants described difficulty in identifying viable starting points for writing. As S1 noted, \textit{``one of the biggest problems people usually have when brainstorming... is having initial ideas''}. Within this context, students described using AI-supported idea elaboration as a way to broaden the space of ideas under consideration and to evaluate the coherence of emerging directions. Rather than treating suggestions as content to adopt, participants framed them as prompts for reflection--helping them assess whether early ideas made sense or consider alternative perspectives they had not previously identified. As S3 also reflected, \textit{``a lot of the stuff... I hadn't even thought about''}.\looseness=-1

Instructors also described the value of early-stage support for expanding students’ argumentative materials. One instructor highlighted the importance of helping students ideate \textit{``potential lines of argument''} and attend not only to supportive claims but also to \textit{``counter claims that I think I need to engage and respond to''} (I2). Hence, in practice, AI-supported idea elaboration could be appropriated to scaffold exploration during pre-drafting, supporting students in articulating and evaluating possible argumentative directions before committing to a draft. This interpretation is consistent with survey responses indicating that students perceived the greatest usefulness during pre-writing and early drafting stages, particularly for brainstorming and structuring arguments (Fig.~\ref{fig:survey}).\looseness=-1

\vspace{-1em}
\subsection{Using Visual Representations to Reason About Argument Structure}

During drafting and revision, students frequently relied on visual representations--such as spatial layouts of claims, evidence, and counterarguments--to reason about the organization and coherence of their arguments. Externalizing argumentative components into visual form allowed students to inspect relationships that were difficult to perceive in raw text. For example, S3 explained that \textit{``the flow chart was really helpful to see where I’m going from each point''}, while S1 noted that visualizing argumentative flow made logical connections easier to identify and revise.

Instructors similarly framed visualization as a classroom scaffold for supporting structural reasoning. One instructor described how visual layouts helped students see the argument as a whole,  making it easier to understand how individual components functioned and related to one another, including \textit{``how the evidence supports the claim [and] how the counterargument offers a different perspective''} (I1). Another instructor emphasized that visual representations supported awareness of \textit{``the connections between ideas and relationships between paragraphs''} (I2). These perspectives suggest that argument visualization functioned as a representational resource for inspecting and revising argumentative structure.
Consistent with these feedbacks, students’ activity during the deployment frequently involved visual argument editing (Tab.~\ref{tab:interaction-distribution}), and survey responses highlighted strong perceived helpfulness for visual editing and viewing argument structure (Fig.~\ref{fig:helpfulness}). In the context of classroom use, these patterns suggest that visual representations supported students’ engagement in structural reasoning by making argumentative relationships explicit and inspectable during writing.\looseness=-1

\vspace{-1em}
\subsection{Engaging in Evaluative and Verification Practices}

Across interviews, students characterized their engagement with AI-supported writing as an interpretive and selective process that required active involvement. Participants emphasized evaluating, revising, or discarding system outputs based on alignment with their intentions, reflecting an active stance toward sensemaking during writing. For instance, S1 described the need to actively filter suggestions to retain what was meaningful
When generated language conflicted with stylistic expectations, students revised or rejected it altogether; as S4 noted, \textit{``this is never something that I would say''}.

From a learning perspective, these interactions reflect forms of active learning and critical engagement, in which students articulated preferences, monitored alignment with goals, and exercised judgment during revision. Instructors highlighted the importance of preserving such engagement, but still noted that introducing AI support too early may constrain opportunities for independent reasoning. One instructor explained that early reliance \textit{``might do way too much of the work for them''} and diminish benefits as thinkers and as writers (I1). Instructors described positioning AI support during reflective stages following initial drafting, where prompts such as \textit{``if you were to rewrite this paragraph, how would you modify it?''} (I1) could scaffold reflection and critical evaluation. This instructional orientation is consistent with interaction logs showing greater engagement in planning and visual revision activities than in draft regeneration (Tab.~\ref{tab:interaction-distribution}).

Students also reported engaging in verification practices when encountering unreliable AI-generated information, illustrating a critical orientation toward AI assistance. Several participants described noticing hallucinated references, including nonexistent authors or invalid URLs, which prompted them to consult external scholarly sources. Rather than disengaging, students interpreted these instances as signals to scrutinize AI-provided content. As S3 remarked, \textit{``you do need to check more carefully,''} reflecting a developing practice of treating AI outputs as tentative and subject to verification. Although survey responses suggest that hallucinated content was not widely perceived as a major challenge (Fig.~\ref{fig:survey}), interview data suggest that in some cases, infrequent inaccuracies still prompted some practices of source evaluation and trust calibration.\looseness=-1

Instructors framed these patterns as aligned with instructional goals of fostering students’ critical understanding of AI’s role and limitations in academic work. As one instructor explained, this emphasis involved \textit{``more an understanding of how AI works''} (I2). These suggested that a less conversational interface may discourage the uncritical acceptance often associated with chat-based systems.\looseness=-1

\vspace{-1em}
\subsection{Metacognitive Engagement}
Across interviews, students described engaging with AI-supported writing in ways that involved monitoring and adjusting their own writing. Students treated AI feedback as a resource for reflecting on intentions, identifying misalignments, and determining revision strategies. As S4 noted, engaging with the system involved \textit{``critical thinking''} and \textit{``sorting through what I want to say''} reflecting sustained attention to their own decision-making during writing.

Several students reported increased awareness of recurring issues through repeated interaction and described using this awareness to guide subsequent decisions. For example, S2 explained that prior feedback helped them anticipate revisions—\textit{``I did X wrong, so I need to do Y''}—even before new prompts appeared. Similarly, S3 observed, \textit{``even if I’m not using VISAR, I can see [the misalignments],''} suggesting attention to structural patterns in their own writing. We interpret these descriptions as evidence of short-term metacognitive awareness, particularly in noticing and evaluating alignment with writing goals.

While the observed patterns suggest metacognitive engagement during writing, the one-week deployment window limits our ability to interpret these findings as evidence of enduring habit formation or long-term learning gains. Instead, we view them as indicative of short-term metacognitive engagement. This interpretation aligns with pre/post quiz results showing short-term conceptual gains in Sec.~2 and survey responses emphasizing improvements related to logic and structure (Fig.~\ref{fig:improvement}). 
At the same time, instructors also suggested that instructional framing and curricular integration might also influence how such metacognitive engagement unfolds, beyond the tool itself. As one instructor noted, effective use \textit{``it would have to be on the instructor to implement it more thoughtfully''} (I1) underscoring that opportunities for reflection and self-monitoring are situated within broader pedagogical practices.
\vspace{-0.2cm}
\section{Discussion}
\subsection{Learning-Oriented AI Writing Support in Authentic Classroom Contexts}

As AI writing assistants become increasingly integrated into students’ learning processes, prior work has raised concerns about potential negative impacts of long-term reliance on large language models at neural, linguistic, and behavioral levels, as well as broader negative educational implications~\cite{kosmyna2025brainchatgptaccumulationcognitive}. Our findings suggest that when AI writing tools embed cognitive scaffolds into the writing process, they can effectively guide students toward engaging with cognition-preserving scaffolded interactions. This indicates that the learning value of AI writing tools in classroom contexts lies in their role as mediating artifacts that support students’ engagement with core writing processes---including ideation, structural reasoning, and revision~\cite{flowerHayes1981cognitive}.
Specifically, students leveraged AI-based idea elaboration and visualizations of argumentative structure (Tab.~\ref{tab:interaction-distribution}; Fig.s~\ref{fig:clusters} and~\ref{fig:helpfulness}) as resources for sensemaking rather than as substitutes for authorial decision-making. These features scaffolded students’ generation, comparison, and evaluation of ideas across writing stages (Sec.~\ref{interaction_patterns}; Fig.~\ref{fig:survey}), enabling them to move from under-specified starting points toward more structured argumentative plans and to surface structural issues that were previously implicit in raw text during reflection.

Crucially, this pattern reflects a form of learning-oriented appropriation: when AI support is embedded in process-oriented writing activities, students tend to engage with AI as a resource for planning and reflection rather than as a shortcut for content production~\cite{bhat2023interacting, su2023collaborating}. This interpretation is further supported by our quantitative findings, which show positive associations between targeted scaffold use (e.g., Spark Usage) and multiple dimensions of argument quality (Tab.~\ref{tab:reg_quality}). These results indicate that cognition-preserving affordances can meaningfully influence how students integrate AI into their writing workflows, supporting engagement in cognitively demanding practices that underlie argumentative learning.
\looseness=-1

\vspace{-1em}
\subsection{Metacognitive Regulation Across Engagement Patterns}

We conceptualize argumentative writing as a task that requires metacognitive control—that is, learners’ ability to monitor, evaluate, and strategically regulate how claims, reasons, and evidence are generated and coordinated~\cite{schrawDennison1994MAI, zimmerman2002srl}. Our interaction-based clustering analysis suggests that AI-supported writing tools can support this form of regulation, although it manifests differently across engagement patterns (Fig.~\ref{fig:clusters}; Sec.~\ref{interaction_patterns}).\looseness=-1

Students categorized as \textit{Strategic Scaffolders} exhibited interaction traces consistent with sustained metacognitive regulation. Their writing processes reflected an integrated workflow where students repeatedly externalized argumentative structure through the visual workspace, invoked targeted scaffolds (e.g., prompts related to evidence), and iteratively revised their drafts. This pattern aligns with metacognitive monitoring (checking for gaps or inconsistencies) and metacognitive control (deciding what to revise next), rather than with unstructured expansion of the idea space~\cite{flavell1979metacognition, schraw1998promoting}. Consistent with this, regression analyses indicate that \textit{Spark Usage} and \textit{Visual Planning Actions} are positively associated with dimensions of argument quality (Tab.~\ref{tab:reg_quality}). \looseness=-1

In contrast, \textit{Brainstorm-Oriented Explorers} interacted more with idea elaboration to explore possibilities. This pattern is more consistent with metacognitive activity oriented toward planning and idea generation, but with fewer observable instances of sustained evaluation and revision within the tool. For \textit{Visual Planners}, interaction traces were dominated by intensive visual restructuring with comparatively limited use of other features, suggesting a preference for externalizing and refining argument structure through spatial representations. Importantly, these engagement patterns likely reflect different emphases in metacognitive regulation under varying writing stages or instructional conditions~\cite{hayes2012modeling}.

Since our clustering analysis is based on interaction traces rather than direct measures of metacognition, we treat these differences as suggestive rather than definitive. Future work could triangulate these archetypes using think-aloud protocols, validated metacognitive instruments, or finer-grained indicators of revision cycles to more precisely examine the relationship between scaffold use and metacognitive control.\looseness=-1

\vspace{-1em}
\subsection{Critical AI Literacy as Situated Practices}
Critical AI literacy broadly refers to people’s ability to critically interpret, evaluate, and make informed judgments about AI-generated outputs, including an awareness of system limitations, potential errors, and appropriate degrees of trust~\cite{leander2020critical, huang2023using, kasneci2023chatgpt, tang2024developer}. In this work, we conceptualize critical AI literacy not as a general disposition, but as a situated capability that is enacted through concrete writing practices with AI-supported writing tools.

A recurring theme across interviews was that students treated AI output as negotiable rather than authoritative, enacting critical AI literacy through situated writing activities. Students reported encountering hallucinated citations or unreliable information and responded by verifying claims using external databases. These verification-oriented behaviors illustrate how structured writing tasks can create opportunities for students to practice working effectively with AI, embedding an understanding of system limitations directly within task performance. In this sense, critical AI literacy emerges not as a standalone skill, but as a form of situated practice that arises from the demands and affordances of the writing activity.\looseness=-1

Our findings also suggest that interface designs emphasizing structure, inspection, and iterative revision—rather than one-shot answer generation—could be helpful to supporting students’ decision-making roles by positioning AI output as material to be evaluated and transformed. However, concerns remain that even within such designs, reliance on generative AI may still lead to cognitive offloading and reduced engagement~\cite{kosmyna2025your, zare2025chatgptEssayRecall, jin2025agencyGap, tang2025naturaledit, tang2025exploring}. Future work could further investigate these dynamics through longitudinal classroom studies, examining how sustained exposure to different AI-supported writing designs influence students’ critical AI literacy over time. \looseness=-1

\vspace{-0.5em}
\subsection{Instructional Mediation and Classroom Integration}

Our results indicate that learning outcomes associated with AI-supported writing tools are jointly influenced by instructional framing, classroom integration, and the design of the AI-assisted writing tool itself. Instructors highlighted the uncertainty inherent in classroom deployment, including when to introduce the tool, how long students should engage with it, and how it should be integrated into instruction. Given limited class time, they described having to make deliberate trade-offs between introducing the tool, allocating time for student interaction, and attending to other instructional activities.
As a result, the same tool may result in markedly different pedagogical outcomes depending on how it is introduced and integrated. In our deployment, instructors introduced the AI-supported writing tool only after students had produced an initial draft, and the observed engagement patterns and learning outcomes may therefore differ from scenarios in which such tools are introduced earlier in the writing process. More broadly, learning opportunities were not automatically ``produced'' by AI, but emerged through intentional pedagogical framing and integration within instructional contexts. \looseness=-1

From this perspective, classroom deployment should be understood as a sociotechnical intervention whose outcomes arise from interactions among tool affordances, instructional decisions, classroom norms, and assessment practices. Educational tools that appear effective in one course may function quite differently once embedded within another course environment.
Accordingly, future work evaluating AI-supported writing tools could move beyond assessing whether such tools help in one course and instead examine how the same tools are used across different classes by incorporating measures of instructional conditions. \looseness=-1

\vspace{-0.5em}
\subsection{Limitations and Future Work}

Our classroom deployment lasted one week. Therefore, our findings should be interpreted as evidence of short-term uptake and immediate impact rather than long-term skill development or enduring habit formation. Although we observed short-term learning gains and self-reported shifts in how students approached argument planning and revision, we lack longitudinal data to assess retention or transfer across subsequent assignments.

In addition, while visual representations supported structural reflection, they also introduced usability challenges as argument complexity increased. Some students described densely populated visual maps as overwhelming, suggesting that increasing structural visibility need to be balanced against cognitive load. Future work could explore adaptive visualization techniques or progressive disclosure to better support complex argumentation contexts.

Accordingly, future research could adopt longitudinal, multi-assignment designs to examine how different exposure, assignment design, and instructional guidance jointly influence sustained uptake. This could involve tracking the same students across multiple writing tasks with delayed post-tests to assess retention, experimentally varying instructional sequencing (e.g., allowing AI use before a human-authored outline), and evaluating whether early improvements in writing persist once AI support is withdrawn.
\section{Conclusion}
This paper examined how an AI-supported writing tool can support student argumentative writing leaning in real classroom settings. Through a combination of quantitative and qualitative analyses, we identified distinct patterns in how students balanced visual planning, AI-assisted ideation, and manual drafting, and evaluated short-term learning outcomes related to argumentative writing. Our results provide empirical evidence on how AI-based writing scaffolds can support students’ argumentative writing learning, offering insights for the design of future learning-oriented AI writing tools.\looseness=-1

\balance

\bibliographystyle{ACM-Reference-Format}
\bibliography{ref}
\end{document}